\documentclass[twocolumn,showpacs,showkeys,preprintnumbers,amsmath,amssymb]{revtex4}

\usepackage{graphicx}
\usepackage{dcolumn}
\usepackage{bm}

\begin{document}

\preprint{Phys.Rev.Letters}

\title{Quantum Hall effect near the charge neutrality point in
two-dimensional electron-hole system}
\author{G.M.Gusev,$^1$ E.B.Olshanetsky,$^{1,2}$ Z.D.Kvon,$^2$ N.N.Mikhailov,$^2$
S.A.Dvoretsky,$^{2}$ and J. C. Portal$^{3,4,5}$}

\affiliation{$^1$Instituto de F\'{\i}sica da Universidade de S\~ao
Paulo, 135960-170, S\~ao Paulo, SP, Brazil}
\affiliation{$^2$Institute of Semiconductor Physics, Novosibirsk
630090, Russia} \affiliation{$^3$LNCMI-CNRS, UPR 3228, BP 166, 38042
Grenoble Cedex 9, France} \affiliation{$^4$INSA Toulouse, 31077
Toulouse Cedex 4, France} \affiliation{$^5$Institut Universitaire de
France, 75005 Paris, France}

\date{\today}
\begin{abstract}
We study the transport properties of $HgTe$-based quantum wells
containing simultaneously electrons and holes in magnetic field B.
At the charge neutrality point (CNP) with nearly equal electron and
hole densities, the resistance is found to increase very strongly
with B while the Hall resistivity turns to zero. This behavior
results in a wide plateau in the Hall conductivity
$\sigma_{xy}\approx 0$ and in a minimum of diagonal conductivity
$\sigma_{xx}$ at $\nu=\nu_p-\nu_n=0$, where $\nu_n$ and $\nu_p$ are
the electron and hole Landau filling factors. We suggest that the
transport at the CNP point is determined by electron-hole "snake
states" propagating along the $\nu=0$ lines. Our observations are
qualitatively similar to the quantum Hall effect in graphene as well
as to the transport in random magnetic field with zero mean value.

\pacs{71.30.+h, 73.40.Qv}

\end{abstract}

\maketitle

The quantum Hall effect (QHE) of a two-dimensional (2D) electron
gas in a strong magnetic field is one of the most fascinating
quantum phenomena discovered in the condensed matter physics. Its
basic experimental manifestation is a vanishing longitudinal
conductivity $\sigma_{xx}\approx 0$ and a quantization of the Hall
conductivity $\sigma_{xy}=\nu\frac{e^{2}}{h}$, where $\nu$ is the
Landau filling factor \cite{sarma}. The discovery of a 2D
electron-hole system in graphene at a finite magnetic field has
put a beginning to a series of studies of the properties of the
special state realized when the densities of the electrons and
holes are equal, the charge neutrality point (CNP)
\cite{jiang,abanin,checkelsky}. In a strong magnetic field the QHE
near the CNP reveals a plateau in $\sigma_{xy}$ with $\nu=0$
currently associated with the resolution of the spin or the valley
splitting of the lowest Landau level (LL)
\cite{abanin,checkelsky}. However, the longitudinal resistivity
$\rho_{xx}$ at $\nu=0$ demonstrates different behavior in various
but otherwise quite similar samples: in some samples $\rho_{xx}$
shows a rapid divergence at a critical field and, with the
temperature decreasing, saturates at a value much larger than the
quantum of resistance \cite{checkelsky}, while in the others it
decreases with lowering the temperature \cite{abanin}. The QHE
behavior near the CNP has attracted much theoretical interest and
several microscopic mechanisms that might be responsible for these
phenomena have been proposed \cite{abanin,shimshoni}, however no
final quantitative conclusion has been drawn up yet.\\
Graphene remained a unique 2D system with described properties
when recently a new 2D system showing similar properties has been
discovered. It has been shown \cite{kvon} that a two-dimensional
semimetal exists in undoped HgTe-based quantum wells with an
inverse band
structure and a (013) surface orientation.\\
In this paper we report the results of our study of the transport
properties of the HgTe-based quantum wells near the CNP. At filling
factor $\nu=0$ (CNP) our system goes into a high resistivity state
with a moderate temperature dependence markedly different from a
thermally activated behavior expected when there is an opening of
the Landau gap in the density of states. We suggest that at the CNP
the 2D electron-hole gas in our HgTe quantum wells is not
homogeneous due to the random potential of impurities. The random
potential fluctuations induce smooth fluctuations in the local
filling factor around $\nu=0$. In this case the transport is
determined by special class of trajectories, the "snake states"
\cite{muller} propagating along the contours $\nu=0$. The situation
is very similar to the transport of two-dimensional particles moving
in a spatially modulated random magnetic field with zero mean value
\cite{aronov}. We especially emphasize that our results may be
equally relevant to the composite fermions description of the
half-filled Landau level \cite{heinonen} and quantum Hall effect in
graphene at Dirac point \cite{jiang,abanin,checkelsky}. \\
\begin{figure}[ht!]
\includegraphics[width=9cm,clip=]{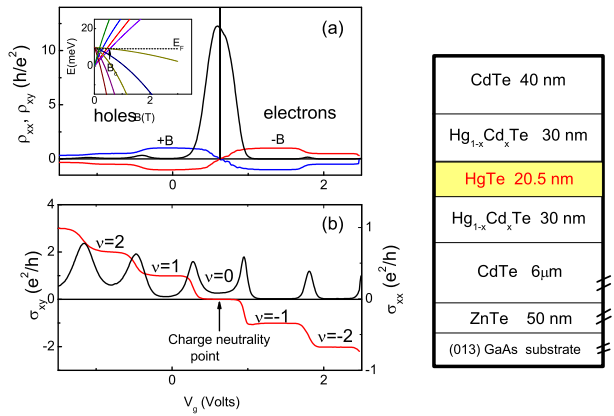}
\caption{\label{fig.1}(Color online) Top-schematic view of the sample.
(a) Diagonal $\rho_{xx}$
(black) and Hall $\rho_{xy}$ (blue, red) resistivities as a
function of the gate voltage at fixed magnetic field B=2.8 T. Hall
resistivity is shown for different signs of the magnetic field.
Insert -Landau level fan diagram for the electron and hole
subbands. Notice that the last Landau levels from each set of
subband cross at a finite $B_{c}$-field. $E_{F}$ is the Fermi
level at the charge neutrality point. (b) Diagonal $\sigma_{xx}$
(black, right axis) and Hall $\sigma_{xy}$ (red, left axis)
conductivities as a function of the gate voltage at the same
magnetic field, T=90 mK. Arrow indicates the position of the
charge neutrality point, when $n_{s}=p_{s}$. }
\end{figure}
Our HgTe quantum wells were realized on the basis of undoped
CdHgTe/HgTe/CdHgTe heterostructure grown by means of MBE at
T=160-200 C on GaAs substrate with the (013) surface orientation.
The details of the growth conditions are published in
\cite{mikhailov}. The section of the structure under investigation
is schematically shown in Fig. 1. For magnetotransport
measurements $50 \mu$m by $100 \mu$m Hall bar samples have been
fabricated on top of these quantum wells by standard
photolithography. The ohmic contacts to the two-dimensional gas
were formed by the in-burning of indium. To prepare the gate, a
dielectric layer containing 100 nm $SiO_{2}$ and 200 nm
$Si_{3}Ni_{4}$ was first grown on the structure using the
plasmochemical method. Then, the TiAu gate was deposited. The
density variation with gate voltage was $8.7\times 10^{14}
m^{-2}V^{-1}$. The magnetotransport measurements in the described
structures were performed in the temperature range 0.050-4.1 K and
in magnetic fields up to 7 T using a standard four point circuit
with a 2-3 Hz ac current of 0.1-1 nA through the sample, which is
sufficiently low to avoid the overheating effects. Several samples
from the same
wafer have been studied.\\
When a large positive voltage is applied to the gate, the usual
increase of the electron density is observed. At lower gate
voltages there is a coexistence of electrons and holes with close
densities \cite{kvon}. Finally, for large negative voltages the
Fermi level goes deep into the valence band and the sample becomes
p-conductive. The density of the carriers at CNP without magnetic
field was $n_{s}=p_{s}\approx 5\times10^{10} cm^{-2}$, the
mobility corresponding mobility was $\mu_{n}=250000 cm^{2}/Vs$ for
electrons and $\mu_{p}\approx25000 cm^{2}/Vs$ holes. These
parameters were found from comparison of the Hall and the
longitudinal magnetoresistance traces with the Drude theory for
transport in the presence of two types of carriers \cite{kvon}. In
magnetic field the energy spectrum of electrons and holes is
quantized and, naively, the LL fan diagram consists of two sets of
overlapped LLs, as shown in figure 1. Above some critical magnetic
field $B_{c}$ it is expected that a zeroth LL gap will open in the
spectrum after the last hole and electron LLs cross each
other.\\
Figure 1  (a) shows the longitudinal $\rho_{xx}$ and Hall
$\rho_{xy}$ resistivities as a function of the gate voltage at fixed
magnetic field. Pronounced plateaux with values $\rho_{xy}=-h/\nu
e^{2}$ are clearly seen at $\nu=-2,-1,1,2$ accompanied by deep
minima in $\rho_{xx}$ on electron and hole sides of the dependence
(here $\nu=\nu_p-\nu_n$ -is the differential filling factor).
Surprisingly, when $V_{g}$ is swept through the charge neutrality
point the longitudinal resistivity shows a large maximum, whereas
$\rho_{xy}$ goes gradually through zero from $h/e^{2}$ ($-h/e^{2}$)
value on the electron side to $-h/e^{2}$ ($h/e^{2}$) value on the
hole side. Indeed, $\rho_{xx}$ is symmetric, and $\rho_{xy}$ is
antisymmetric in B. Fig.1b shows the conductivities $\sigma_{xx}$
and $\sigma_{xy}$ as a function of the gate voltage calculated from
experimentally measured $\rho_{xx}$ and $\rho_{xy}$ by tensor
inversion. Standard quantum Hall effect plateaux
$\sigma_{xy}=e^{2}\nu/h$ accompanied by minima in $\sigma_{xx}$ are
clearly visible. We note, however, that the steps in $\sigma_{xy}$
on the holes side are not completely flat and the minima in
$\sigma_{xx}$ are not very deep due to a lower hole mobility. Notice
that the height of the peaks in $\sigma_{xx}$ is very close to
$e^{2}/2h$ for all Landau levels as expected for conventional QHE.
The most intriguing QHE state is observed at the charge neutrality
point at $\nu=0$. Figure 1b shows that $\sigma_{xy}$ has a flat zero
value quantum Hall plateau around CNP, while $\sigma_{xx}$ displays
a pronounced minimum. Both the minimum and the plateau at CNP
strongly depend on the magnetic field and the temperature. Figure 2
demonstrates the evolution of $\sigma_{xx}$ and $\sigma_{xy}$ with
temperature (a) and magnetic field (b) when the system is driven
from n-type to p-type. Indeed, all QHE states become more pronounced
with T decreasing and B increasing, especially the plateau and the
minimum on hole side closest to the CNP. The position of all the
minima, except the minima at $\nu=0$, shifts with magnetic field, as
a result of the increase in the LLs degeneracy.

\begin{figure}[ht!]
\includegraphics[width=9cm,clip=]{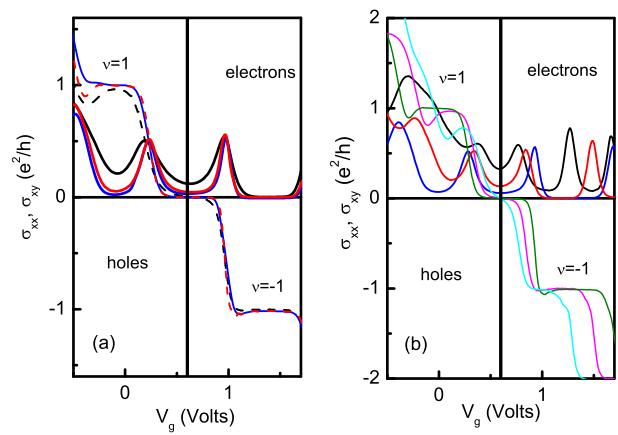}
\caption{\label{fig.2}(Color online) (a) Diagonal
$\sigma_{xx}$(solid lines) and Hall $\sigma_{xy}$(dashed lines)
conductivities as a function of the gate voltage at fixed magnetic
field B=2.8 T and at the different T (mK) : 850( black), 250(blue),
90 (red). (b) Diagonal $\sigma_{xx}$ (solid lines) and Hall
$\sigma_{xy}$ (dashed lines) conductivities as a function of the
gate voltage at the different values of the magnetic field B(T) :
1.5(black), 2(red), 2.5 (blue), T=50 mK. Vertical line indicates
CNP. }
\end{figure}
In the rest of the paper we will focus on the magnetoresistance
behaviour at the CNP. Figure 3 shows a sharp increase in the
resistivity $\rho_{xx}$ with magnetic field at the CNP  at
different temperatures. It is worth noting that all the
resistivity curves show a plateaux-like feature $\rho_{xx}\sim
h/4e^{2}$ visible at $B_{c}\approx 1.4 T$. The resistivity shows
no temperature dependence below $B_{c}$, while above 1.4 T
$\rho_{xx}$ increases the more rapidly the lower the temperature.
Indeed such temperature dependence may indicate activated
behaviour due to the opening of a zeroth gap, as expected from the
simple energy diagram shown in the Figure 1. Surprisingly, we find
that the profile of $\rho(T)$ does not fit the activation form
$\rho(T)\sim exp(\Delta/2kT)$, where $\Delta$ is the activation
gap. Insert to figure 3 shows that $\rho(T)$ starts to deviate
strongly from the Arrenius exponential law and at low temperature
and high magnetic field $\rho_{xx}$ it almost saturates. Therefore
we may conclude that the observed T dependence is not related to
the opening of the gap at the CNP and another mechanism is
responsible for this behaviour.
\begin{figure}[ht!]
\includegraphics[width=9cm,clip=]{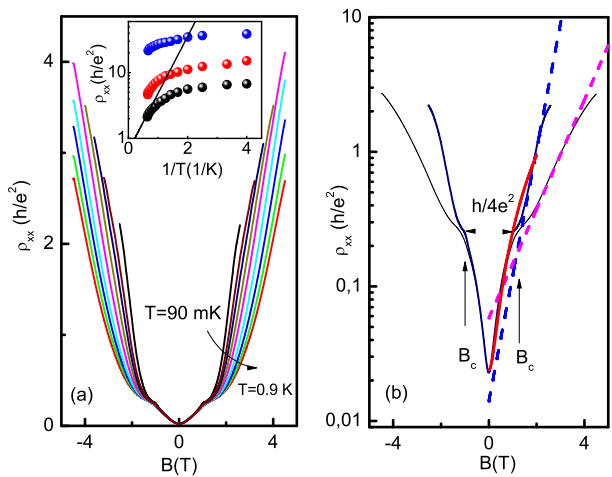}
\caption{\label{fig.3}(Color online) (a) Magnetoresistivity at CNP
for different temperatures. Insert: the resistivity as a function of
the inverse temperature at a fixed magnetic field B(T): 4
(black),6(red), 7(blue). The solid line in the insert is a fit of
the data with an Arrenius function with $\Delta=0.3 meV$ (b)
Magnetoresistivity at CNP for two temperatures (90 and 900 mK).
Magenta and blue dashed straight lines are fits of the data with the
function $\rho(T)\sim exp(\Delta/2kT)$. Red solid curve is the
fitting obtained using the theoretical approximation \cite{evers}
describing snake state percolation in quasiclassical random magnetic
field regime. }
\end{figure}
The important indication of the existence of
the gap at CNP may be the exponential increase of $\rho_{xx}$ with
B. Note that the energy gap at CNP is determined by equation
$\Delta=\frac{\hbar\omega_{c}^{e}}{2}+\frac{\hbar\omega_{c}^{p}}{2}-\Delta_{0}$,
where $\hbar\omega_{c}^{e,p}$ is the cyclotron energy of electron or
hole, $\omega_{c}^{e,p}=\frac{eB}{m_{e,p}c}$ is the cyclotron
frequency, $m_{e,p}$ is effective mass, $\Delta_{0}\approx5 meV$ is
the overlapp of the subbands \cite{kvon}. Taking into account the
effective masses $m_{e}=0.025m_{0}$, $m_{p}=0.15m_{0}$ meV at CNP,
we obtain $\Delta >7.4 meV$ at $B>3 T$.
\begin{figure}[ht!]
\includegraphics[width=9cm,clip=]{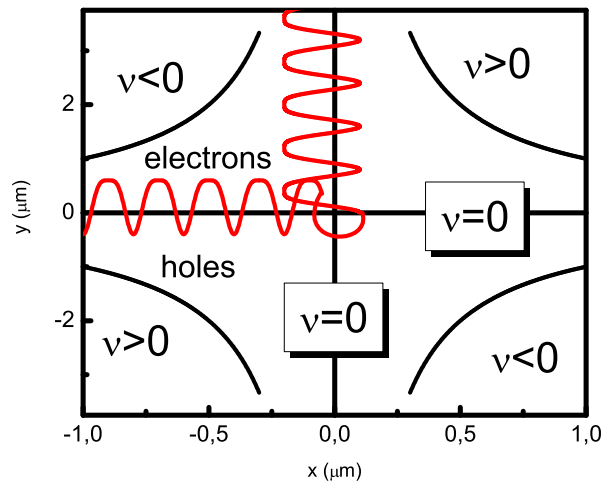}
\caption{\label{fig.4}(Color online)(a) Schematic illustration of
the electron-hole "snake state" percolation along $\nu=0$ line at
CNP in the strong magnetic field and geometry of the saddle point
between adjacent percolation clusters.}
\end{figure}

Figure 3 shows the comparison of the magnetoresistivity at the CNP
with equation $\rho(T)\sim exp(\Delta/2kT)$, taking into account
the linear dependence of the gap $\Delta$ on the magnetic field.
The increase of $\rho_{xx}$ with B is qualitatively consistent
with the opening of the gap, however, the value of $\Delta$ is
found to be smaller by more than a factor of 30. Moreover, we
found the reduction of the deduced gap with lowering the
temperature, which seems very unlikely. Therefore, a closer
inspection of the $\rho_{xx}(B)$ data raises further doubt as to
the existence of a gap
at $\nu=0$ filling factor.\\
To understand this anomalous quantum Hall effect at CNP in the
two-dimensional electron-hole system static disorder should be
taken into account i.e. the fluctuations of the local filling
factor around $\nu=0$ induced by the smooth inhomogeneities.
Notice that at $B
> B_{c}$ the opening of the gap at $\nu=0$ leads to depopulation
of the levels, and the system turns into a conventional insulator
with $n_{s}=p_{s}\rightarrow 0$. From a simple argument this occurs
when
$\frac{\hbar\omega_{c}^{e}}{2}+\frac{\hbar\omega_{c}^{p}}{2}=\Delta_{0}$.
For $\Delta_{0}\approx 5 meV$ we obtain $B_{c}=1.4 T$, which agrees
well with the position of the plateau-like feature in $\rho_{xx}(B)$
dependence, shown in figure 3.\\
Fluctuations of the local filling factor $\nu$ near zero leads to
the formation of percolation paths along the $\nu=0$ contours. A
remarkable point to be noted is the possibility of a conducting
network formed of such contours, since the coupling between two
adjacent percolating cluster occurs through the critical saddle
points, as shown in figure 4 \cite{lee, evers}. The conductivity
is determined by the electrons and holes that move along $\nu=0$
lines and in quasiclassical regime have a trajectories of a snake
type, which is shown in figure 4. We emphasize the similarities in
the description of the conductivity of an electron-hole system at
the CNP and the transport of two-dimensional electrons in a random
magnetic field with zero mean
value \cite{lee, evers, khveschenko}.\\
Propagation of the snake-type trajectories along the $\nu=0$
contours and their scattering at the saddle points may explain the
large magnetoresistance at the CNP in figure 3. The analytical
solutions for the model \cite{evers} have been obtained for two
distinct regimes corresponding to a small and a large amplitude of
the random magnetic field (RMF). In the limit of a small RMF at
$\alpha <<1$, where $\alpha=d/R_{c}$, $d$ is the correlation radius
of the potential fluctuations, $R_{c}=\frac{\hbar k_{F}}{m_{e,p}
\omega_{c}}$ is the Larmour radius, $k_{F}$ is the Fermi wave
number, the conductivity is given by
$\sigma_{xx}=\frac{e^{2}}{h}\frac{k_{F}d}{4\alpha^{2}}$. Weak
disorder regime $\alpha<<1$ in our case corresponds to the regime of
a small amplitude random magnetic field. Figure 3b shows the results
of the comparison of our data and the theory of percolation via the
snake states. We obtain an excellent agreement with parameter
$d=0.06 \mu m$, which seems very reasonable. Note that the classical
Drude model predicts quadratic positive magnetoresistance $\Delta
\rho_{xx}/\rho_{0}=\mu_{n}\mu_{p}B^{2}$ and large positive Hall
resistance $\rho_{xy}=B/ne$ for homogeneous electron-hole system at
CNP in the case $\mu_{n}>>\mu_{p}$ and $n=n_{s}=p_{s}$ \cite{kvon}.
Our magnetoresistance data and observation of the $\rho_{xy}\approx
0$ are inconsistent with this prediction. We attribute such
difference to the inhomogeneity of the e-h system, as we mentioned
above.  The model \cite{evers} also predicts a crossover from weak
to strong RMF at $\sigma_{xx}\sim h/4e^{2}$, which corresponds to
the features indicated in Figure 3b by arrows at $B_{c}$. In a
strong disorder regime $\alpha>>1$ the theory \cite{evers} predicts
that $\sigma_{xx}\sim \alpha^{-1/2}\ln ^{-1/4}\alpha$, while the
model \cite{khveschenko} gives a different result. Not knowing which
approximation is more realistic we find nevertheless that all models
predict a fast increase of resistance with magnetic field in
accordance with our observation. There is also one important
observation to be made concerning the strong magnetic field
approximation. We expect that with magnetic field increasing the
density inhomogeneity domains will be eliminated at a critical
magnetic field when the magnetic field length falls well below the
typical domains size, and the snake states will be suppressed. For
the estimation of the value of such critical magnetic field a more
detailed theory is required. In this regime the transport should be
dominated by variable range hopping between islands with the
different filling factors. Another scenario in the strong magnetic
field would be a joint contribution of the snake states,
counter-propagating trajectories and variable range hopping to the
transport. However no direct comparison with the experiment in order
to distinguish between all theses models would be possible without
further theoretical and experimental works.  \\
Finally we would like to discuss the similarity and the difference
between other bipolar 2D system, such as graphene \cite{jiang,
abanin,checkelsky} and InAs/GaSb system \cite{nicholas}, where the
QHE has also been studied.
 In contrast both to our 2D e-h system in HgTe QW and to graphene the
 InAs/GaSb-based
system is not a semimetal since there is a gap resulting from the
hybridization of in-plane dispersions of electrons in InAs and
holes in GaSb \cite{yang}. The electrons and holes are spatially
separated by the heterostructure interface. Besides, so far there
has been no demonstration of a state with equal electron and hole
densities in InAs/GaSb (an analogue of the CNP in (013) HgTe QW).
All this results in a qualitatively different interpretation of
the insulating behaviour in the quantum Hall regime. In
\cite{nicholas} a model was proposed which incorporates
counter-propagating edge channels, while we suggest that in our
system there are snake states propagating though the bulk of a
disordered potential landscape. In contrast to the bipolar
InAs/GaSb system, graphene shows a certain similarity with our
system as far as the transport properties  at the CNP in the QHE
regime are concerned \cite{abanin,checkelsky}. The main difference
between our system and graphene is that there are neither
electrons nor holes in graphene at the Dirac point in zero
magnetic field, whereas the HgTe QWs are always populated with
both types of carriers. It allows us to study the transport at
$B<B_{c}$ and compare it with the theoretical snake state model
for a small amplitude random magnetic field regime. Above $B_{c}$
the situation becomes very similar to graphene, except that the
nature of the gap at $\nu=0$ may be different.

In conclusion, we have measured quantum Hall effect near the
charge neutrality point in a system which contains electrons and
holes. We have found that the QHE in this system shows a wide
$\nu=0$ plateau in $\sigma_{xy}$ accompanied by a vanishing
diagonal conductivity $\sigma_{xx}\approx 0$. However, we do not
find a thermally activated temperature dependence in the
longitudinal conductivity minima, which would be expected due to
the opening of a gap in the energy spectrum in a conventional QHE.
We attribute such behavior to a percolation of the snake type
trajectories along $\nu=0$ lines. Our observations show that there
is a common underlying physics in such phenomena as the $\nu=0$
quantum Hall effect at the CNP, QHE in graphene at Dirac point and
the
transport in a random magnetic field with zero mean value.\\
A financial support of this work by FAPESP, CNPq (Brazilian
agencies), RFBI (09-02-00467a and
09-02-12291-ofi-m) and RAS programs "Fundamental researches in
nanotechnology and nanomaterials" and "Condensed matter quantum
physics" is acknowledged.

{\it Note added.}-During the preparation of this manuscript we
became aware of a related work on the application of the random
magnetic field model to transport in graphene by das Sarma {\it al}
\cite{dassarma}.

\end{document}